\documentclass[aps,prx,twocolumn,floatfix,longbibliography,nofootinbib,superscriptaddress]{revtex4-2}
\usepackage[utf8]{inputenc}
\usepackage{natbib}
\usepackage{graphicx}
\usepackage{tikz}\usepackage{xcolor}
\usepackage[tbtags]{amsmath}
\usepackage[colorlinks, linkcolor=blue]{hyperref}
\hypersetup{colorlinks,allcolors=blue}
\usepackage{amssymb}
\usepackage{amsmath}
\usepackage{gensymb}
\usepackage{float}
\usepackage{amsmath}
\usepackage{tabularx,graphicx}
\usepackage{epstopdf}
\usepackage{latexsym}
\usepackage{color, colortbl}
\usepackage{psfrag}
\usepackage{bbm}
\usepackage{bm}
\usepackage{titlesec}
\usepackage{dsfont}
\usepackage{feynmp}
\usepackage{slashed}
\usepackage{multirow}
\usepackage[normalem]{ulem}

\newcommand{\bcen}{\begin{center}}
\newcommand{\ecen}{\end{center}}
\newcommand{\btab}{\begin{tabular}}
\newcommand{\etab}{\end{tabular}}
\newcommand{\bdes}{\begin{description}}
\newcommand{\edes}{\end{description}}

\newcommand{\beq}{\begin{equation}}
\newcommand{\eeq}{\end{equation}}
\newcommand{\bea}{\begin{eqnarray}}
\newcommand{\eea}{\end{eqnarray}}

\newcommand{\half}{\frac{1}{2}}
\newcommand{\bary}{\begin{array}}
\newcommand{\eary}{\end{array}}
\newcommand{\benum}{\begin{enumerate}}
\newcommand{\eenum}{\end{enumerate}}
\newcommand{\bitem}{\begin{itemize}}
\newcommand{\eitem}{\end{itemize}}

\newcommand{\bfig}{\begin{figure}}
\newcommand{\efig}{\end{figure}}
\newcommand{\bra}[1]{{\langle #1 |}}
\newcommand{\ket}[1]{| #1 \rangle}

\newcommand{\eqn}[1] {eqn.~(\ref{#1})}

\newcommand{\Fig}[1]{Fig.~\ref{#1}}

\makeatletter

\newcommand{\Rmnum}[1]{\expandafter\@slowromancap\romannumeral #1@}
\makeatother

\begin{document}

\title{Entanglement signatures of a percolating quantum system}

\date{\today}

\author{Subrata Pachhal}
\email{pachhal@iitk.ac.in}
\affiliation{Department of Physics, Indian Institute of Technology Kanpur, Kalyanpur, UP 208016, India}

\author{Adhip Agarwala}
\email{adhip@iitk.ac.in}
\affiliation{Department of Physics, Indian Institute of Technology Kanpur, Kalyanpur, UP 208016, India}

\begin{abstract}
Entanglement measures have emerged as one of the versatile probes to diagnose quantum phases and their transitions. Universal features in them expand their applicability to a range of systems, including those with quenched disorder. In this work, we show that when the underlying lattice has percolation disorder, free fermions at a finite density show interesting entanglement properties due to massively degenerate ground states. We define and calculate appropriate entanglement measures such as typical, annealed, and quenched entanglement entropy in both one and two dimensions, showing they can capture both geometrical aspects and electronic correlations of the percolated quantum system. In particular, while typical and annealed 
entanglement show volume law character directly dependent on the number of zero modes in the system, quenched entanglement is generally area law albeit showing characteristic signatures of the classical percolation transition. Our work presents an exotic interplay between the geometrical properties of a lattice and quantum entanglement in a many-body quantum system.

\end{abstract}

\maketitle

{\it \textbf{Introduction:}} Quantum entanglement \cite{Einstein_PR_1935, Horodecki_RMP_2009} and their measures \cite{Bennett_PRA_1996, Plenio_QIC_2007, Amico_RMP_2008} have established themselves as necessary tools to diagnose quantum phases and their transitions \cite{Osborne_PRA_2002, Osterloh_Nature_2002, Vidal_PRL_2003,  Refael_PRL_2004, Calabrese_JSTAT_2004, Hamma_PRA_2015, Gioev_PRL_2006, Barthel_PRA_2006, Hastings_JSTAT_2007, Shem_PRL_2013, Canella_SR_2019}. Moreover, the generalizations and extensions of bipartite entanglement entropy (EE)  \cite{Bennett_PRA_1996} including topological EE \cite{Kitaev_PRL_2006, Levin_PRL_2006}, entanglement negativity \cite{Horodecki_PLA_1996, Vidal_PRA_2002, Shapourian_PRB_2017, Shapourian_SP_2019, Shapourian_PRX_2021, Parez_PRR_2024}, witnesses \cite{Lewenstein_PRA_2001, Terhal_TCS_2002, Marko_JPM_2007, Ferenc_PRR_2023} and corner contributions \cite{Fradkin_PRL_2006, Casini_NPB_2007, Bueno_PRL_2015, Helmes_PRB_2016, Jonathan_PRL_2024} provide an encompassing framework to study quantum phases in and out of equilibrium. Their universal features at times make them indispensable to  define unconventional quantum phases \cite{Li_PRL_2008, Fidkowski_PRL_2010, Turner_PRB_2011, Cho_SR_2017, Mondal_PRB_2022}, particularly in the presence of disorder \cite{Prodan_PRL_2010, Liu_PRL_2016}. 
Among disordered systems, percolation problems \cite{hammersley_1957, Essam_IOP_1980, stauffer_1992}, where a lattice is probabilistically diluted either on the sites or bonds, are known to exhibit second-order phase transitions, namely geometrical phase transition, with universal critical exponents \cite{priest_1976, reynolds_1977, Kesten_CMP_1987, stauffer_1992,  Sahimi_Springer_2009}.
For instance, in the case of bond percolation if $p$ is the probability of having a bond in the system such that $p=1$ is a translationally invariant lattice then there exists a critical value of $p$ known as the classical percolation threshold $p_c$, immediately below which the lattice gets geometrically disconnected \cite{stauffer_1992}. Such geometrical phase transitions and their critical phenomena have been long studied in both classical and quantum systems \cite{balian1979ill, fisher_prl1992, fisher_physica99,dutta_prb2002, Chakrabarti_Springer_2009, Kovacs_PRR_22} and recently in the context of topological phases \cite{Sahlberg_PRR_2020, Ivaki_PRR_2020, Mondal_PRBL_2023}. However, the interplay of percolation disorder and entanglement properties has been little explored \cite{Gong_PRB_2009}. We visit entanglement measures in the light of percolation disorder in the simplest of fermionic quantum systems: free fermions hopping on a lattice. In particular, we address do entanglement measures show signatures of a geometrical phase transition? Do they still follow the usual area/volume law \cite{Eisert_RMP_2010, Li_PRB_2006} diagnostics?  

In this work, we investigate the above questions to show that in percolating quantum systems, the conventional measures of bipartite entanglement entropy \cite{Ingo_JPA_2003, Peschel_JPA_2009} have quite a few subtleties. The non-intuitive aspect of the results arises from massive exact degeneracy due to lattice percolation. In particular, we show that one needs to investigate different quantities: {\it typical} ($S_{\text{typ}}$), {\it annealed}  ($S_{\text{ann}}$) and {\it quenched} EE ($S_{\text{quen}}$) each of which captures distinctive signatures. Interestingly, they often depend on the number of exact zero-modes ($N_0$) in the system, which is, in turn, related to the geometrical aspects of the lattice. We finally show that even the classical percolation threshold has footprints in quantum entanglement, where entanglement scaling depends on the emergent fractal nature of the largest cluster.

{\it \textbf{Measures of entanglement and free fermions:}} Given a wavefunction $|\psi \rangle$, or the density matrix ($\hat{\rho} =|\psi\rangle\langle \psi|$) of a system composed of two subsystems $A$ and $B$, the EE of region $A$ with $B$ (or vice-versa) is given by 
$S_{A} = - \text{Tr} \big( {\hat{\rho}_A \ln \hat{\rho}_A} \big)$,
where $\hat{\rho}_A = \text{Tr}_{B} \hat{\rho}$ is the reduced density matrix of the subsystem $A$. However, this definition assumes that the system is described by a unique wavefunction $|\psi\rangle$. But in the presence of degeneracies, it is more appropriate to study {\it typical} entanglement 
\beq
S_{\text{typ}} = \langle S_A(\hat{\rho}_A) \rangle, 
\eeq
where the averaging is done over {\it pure} state ensembles made of degenerate wavefunctions \cite{Page_PRL_1993, Bianchi_PRX_2022}. The wavefunction coefficients are complex and drawn from normal probability distributions implementing a uniform Haar measure over the unitary transformations about any reference state \cite{Bengtsson_Book_2006, Nadal_JSP_2011, Dahlsten_JPA_2014, Biswas_NJP_2021}.  $S_{\text{typ}}$ is however different from 
\beq
S_{\text{ann}} =  S_A(\langle \hat{\rho}_A\rangle),  
\eeq
where the averaging is done on the density matrix before evaluating its entanglement content.
Such a measure we call {\it annealed} EE. In general $S_{\text{typ}}< S_{\text{ann}}$, since the latter reflects the maximum information content possible in a given Hilbert space consistent with Page's result \cite{Page_PRL_1993}. For instance, in a decoupled two spin system while $S_A$ may seem to be zero, $S_{\text{typ}} \sim 0.33$, $S_{\text{ann}} = \ln 2$ (see Supplemental Material (SM) \cite{supp}).  

In a many-body free fermionic system defined on $N$ sites such that $c^\dagger_i$ and $c_i$ are the fermionic creation and annihilation operators, the ground state is often a unique Slater determinant and the bipartite entanglement is evaluated using the Peschel formulation of correlator matrix ${\cal C}$ where its elements $c_{ij}=  \langle c^\dagger_i c_j\rangle$ \cite{Ingo_JPA_2003, Peschel_JPA_2009}. However, the measure of typical and annealed entanglement becomes important for a system with ground state degeneracy. 
Thus, while for any single Slater determinant, the EE can be given by the corresponding correlator matrix $S({\cal C}_A)$, when averaged over all choices of random wavefunction coefficients over the degenerate manifold, one gets $
S_{\text{typ}} = \langle S({\cal C}_A) \rangle$ 
and similarly $S_{\text{ann}} =  S(\langle {\cal C}_A \rangle)$  (for an illustrative example of these quantities in a four-site toy model, see SM \cite{supp}). 
A simple way to break such exact degeneracies would be to add a small perturbative Anderson disorder $\eta \sim 10^{-12}$, which will choose a unique ground state. Then we define {\it quenched} entanglement, to capture the fermionic correlations in the system
\beq
S_{\text{quen}} =  \langle S( {\cal C}_A )\rangle _\eta,   
\eeq
where the average is over disordered configurations.  Before delving into percolation problems, we illustrate the role of these measures in a zero-dimensional system.

\begin{figure}
\centering
    \includegraphics[width=0.99\linewidth]{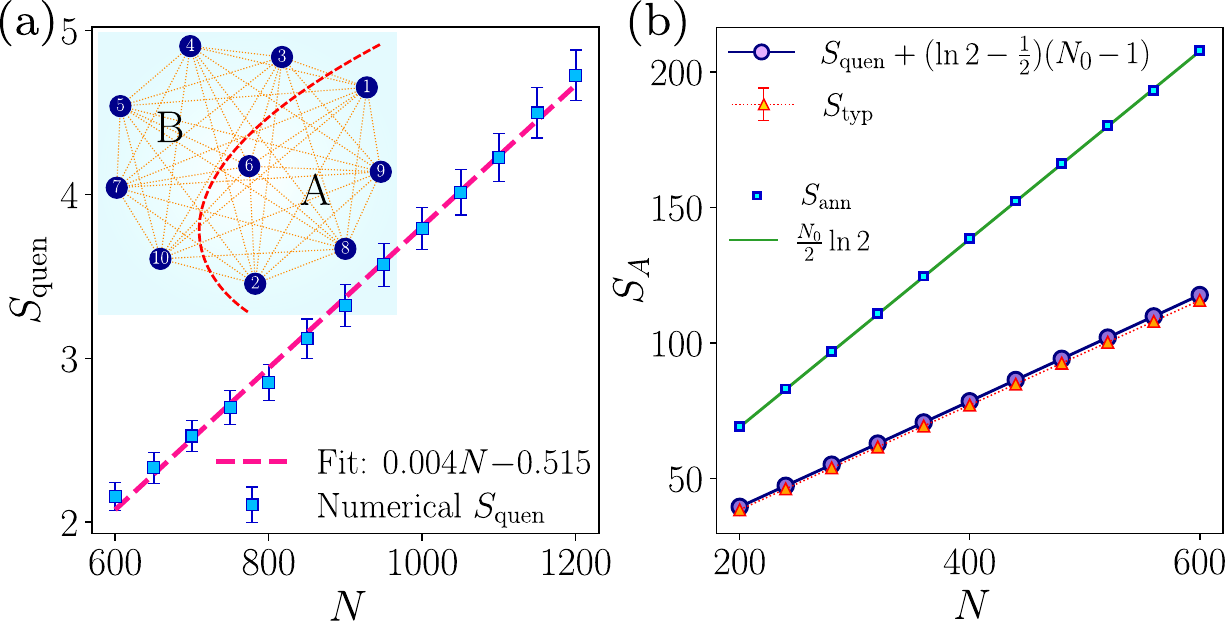}
	\caption{\textbf{Zero dimensional model:} (a)  $S_{\text{quen}}$ (averaged over $N_C\equiv 10^3$ configurations) grows linearly with number of sites, $N$. (inset)  Schematic of the model and subsystems $A$ and $B$. (b) Behaviour of $S_{\text{typ}}$ and $S_{\text{ann}}$ with $N$.  The number of zero-modes, $N_0=N-1$. Both are calculated using $N_R \equiv 10^4$ ensembles of random pure states.}
	\label{fig1}
\end{figure}

{\it \textbf{Zero-dimensional system:}} Consider a $N$ site Hamiltonian where the fermions can hop from a site to any other site with strength $t=-1$ (see \Fig{fig1}(a) inset) such that $H = -\sum_{i,j} c^\dagger_i c_j$. The single particle spectrum has one eigenvalue with energy $-N$, while $N-1$ eigenvalues are exactly $zero$. A half-filled system here is thus ${N-1} \choose {\frac{N}{2}-1}$ fold degenerate. An infinitesimal $\eta$ disorder of strength $\eta=10^{-12}$ can be added in form of a term $\sum_{i}\epsilon_i c^\dagger_i c_i$ where $\epsilon \in [-\eta,\eta]$ to break this exact degeneracy. Given the long-range character of hopping, the entanglement content for equal bipartition of the system such that $N_A = N/2$ is still volume law where $S_{\text{quen}} = s_v N+ s_0$ with a coefficient $s_v \sim 0.004$ as shown in \Fig{fig1}(a). However, when such massive degeneracy is intact, reflects in 
\beq
S_{\text{typ}} = S_{\text{quen}} + (\ln 2 -\frac{1}{2}) (N_{0} -1),
\eeq
where the second term arises from the effective geometrical component because of $N_0=N-1$ number of zero-modes of the system (see \Fig{fig1}(b)). The factor of $(\ln 2 -\frac{1}{2})$ arises from the effective random pure states made out of $N_0$ zero-modes \cite{Liu_PRB_2023}. In general, when the number of occupied zero-modes is $fN_0$ ($0<f<1$), the geometrical volume law is empirically proportional to $f(1-f)$ representing the effective phase space volume (see SM \cite{supp}). Another alternate measure of the entanglement in this system is to average the correlator matrix first. For such a system at half-filling, while $c_{ii}=\frac{1}{2}$ and $c_{ij}=\frac{1}{2N}$ \cite{Gori_PRB_2015} leading to $S_{\text{ann}} \sim \frac{N_{0}}{2} \ln2$.

While it may appear that all three entanglement measures are volume law and therefore similar, it is pertinent to emphasize that they all have different physical content. While $S_{\text{quen}}$ captures the fermionic correlations, which are inherently volume law since the network is zero-dimensional, the character of $S_{\text{typ}}$ and $S_{\text{ann}}$ capture the massive degeneracy of the system - where $S_{\text{typ}}$ measures the average bipartite EE for any choice of a typical {\it pure} state, $S_{\text{ann}}$ measures the maximal subsystem entanglement when the complete system itself becomes {\it mixed} due to averaging of the correlator matrix.  For instance, the entanglement measure of the full system has $S_{\text{ann}} \neq 0$ but $S_{\text{typ}}=0$. In general, when a complete system of interest takes a mixed character various other entanglement measures have been found to isolate quantum correlations between its partitions such as mutual information \cite{Adami_PRA_1997, Berry_PRA_2005, Wolf_PRL_2008}, entanglement negativity \cite{Horodecki_PLA_1996, Vidal_PRA_2002, Shapourian_PRB_2017, Shapourian_SP_2019, Shapourian_PRX_2021, Parez_PRR_2024} and witnesses \cite{Lewenstein_PRA_2001, Terhal_TCS_2002, Marko_JPM_2007, Ferenc_PRR_2023}. Having discussed the subtleties and the different measures of entanglement, we now discuss quantum percolation problems, where studying these various measures of entanglement becomes indispensable given spectral degeneracies.

\begin{figure}
\centering\includegraphics[width=0.99\linewidth]{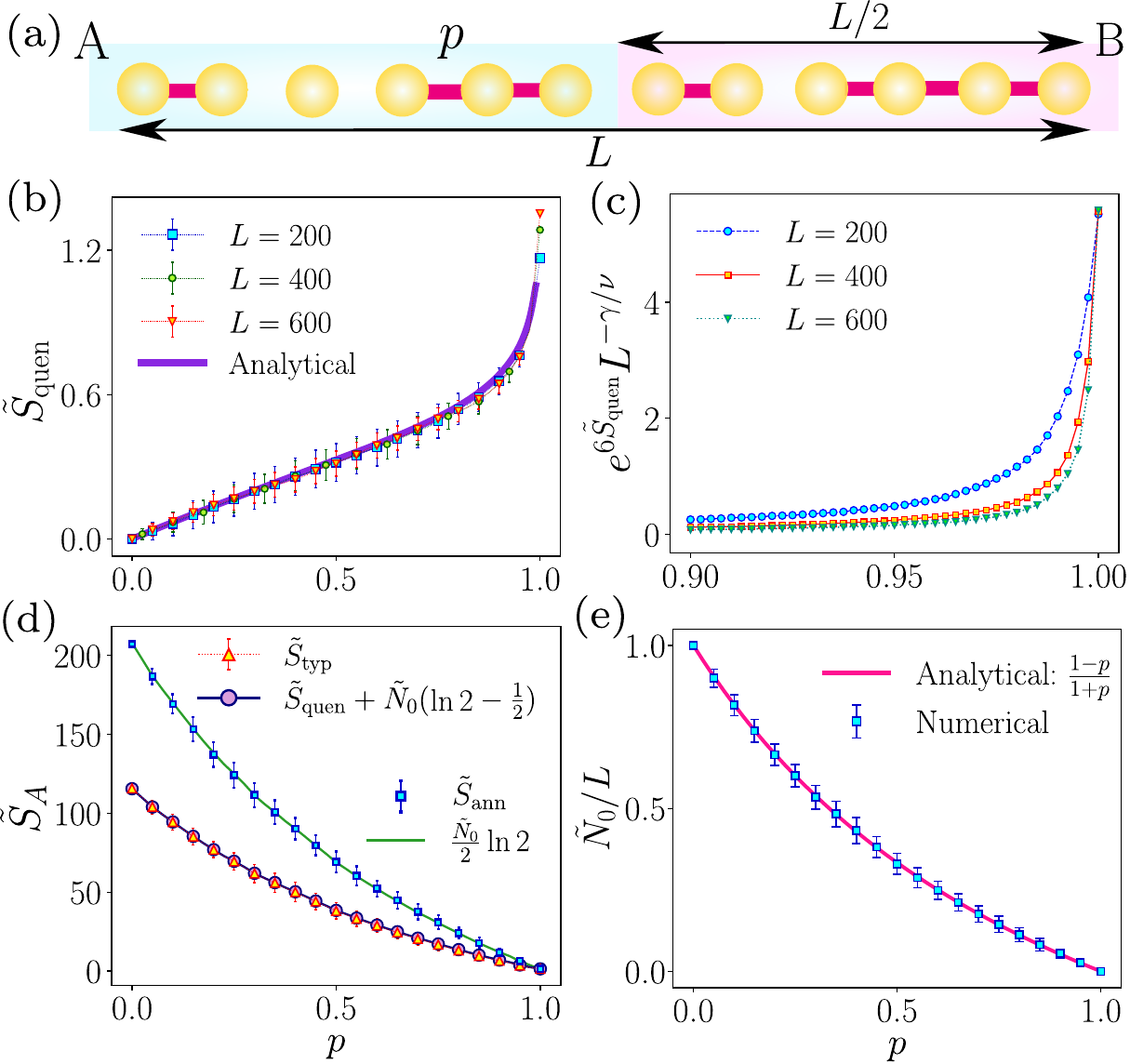}
	\caption{\textbf{Bond diluted chain:} (a) Subsystems $A$ and $B$ in a $L$ sized chain, with bond occupation probability $p$. (b) $\tilde{S}_{\text{quen}}$ with $p$ for different values of $L$ (c) Scaling of $\tilde{S}_{\text{quen}}$ near $p=p_c=1$ with $\gamma=\nu=1$. (d) Behaviuor of $\tilde{S}_{\text{typ}}$ and $\tilde{S}_{\text{ann}}$ with $p$ (e) Zero-mode density ($\equiv \tilde{N}_o/L$) for $L=600$ compared to analytical result \cite{tol}. In (b), (c), and (e) $N_C=10^3$, in (d)  $N_C=40$ and for each configuration, $N_R=10^2$.}
\label{fig2} 
\end{figure}

{\it \textbf{One-dimensional percolation:}} We first discuss percolation in a one-dimensional lattice, where spinless fermions hop with the following tight-binding Hamiltonian
\beq
H = -t \sum_{i=1}^L \Big( c^\dagger_i c_{i+1} + \text{h.c.} \Big),
\eeq
where $L$ is the system size and $t=1$. The probability of having a bond is given by $p$ such that at $p=0$, the system contains a completely decoupled set of sites, while at $p=1$ it is a translationally invariant fermionic chain (see \Fig{fig2}(a)). The percolation transition, where one end of the lattice gets connected to the other end, happens at $p=p_c=1$ \cite{reynolds_1977, Reynolds_IOP_1977}.  The fermion filling is kept fixed at $=1/2$. Given the Hamiltonian is real and has a sublattice symmetry, it belongs to the BDI symmetry class \cite{altland_1997, Agarwala_AOP_2017}, which is retained under the percolation protocol.

The fermionic ground state describes a Fermi sea at $p=1$; however, at any $p<1$, given the fermions reside on disconnected clusters, the ground state should be interpreted as an Anderson insulator state. This is consistent with the effect of any uncorrelated disorder in one dimension\cite{Anderson_PR_1958, Abrahams_PRL_1979}. The entanglement content, therefore, is generically expected to be $\sim \ln L$ at $p=1$ and $\mathcal{O}(1)$ in presence of disorder, as is known from Cardy-Calabrese result for critical states \cite{Calabrese_JSTAT_2004, Wolf_PRL_2006, Gioev_PRL_2006} and area law entanglement for short-range correlated states \cite{Eisert_RMP_2010, Hastings_JSTAT_2007}. 
At any $p$ given the presence of disconnected clusters, there are spectral degeneracies that can be split using an infinitesimal disorder $\eta$ to obtain $S_{\text{quen}}$ (see \Fig{fig2}(b)). An analytical estimate can be obtained as follows. Given $P_s$ is the probability of having a $s$ sized cluster, in general, its maximal entanglement content in equal bipartition is \big($\frac{c}{6} \ln(\frac{s}{\pi}) + c_0$\big) \cite{Calabrese_JSTAT_2004} where $c$ is the central charge ($c=1$) of the one-dimensional bosonic CFT and $c_0$ is an area law piece. Thus for a thermodynamic system,
\beq
\tilde{S}_{\text{quen}} = \sum_{s=2}^{\infty} P_s \Big( \frac{c}{6} \ln(\frac{s}{\pi}) + c_0 \Big)
\eeq
where $\tilde{S}_{\text{quen}}$ represents a configuration averaged value over $S_{\text{quen}}$. 
Here $P_s = s n_{s} = s(1-p)^2 p^{s-1}$
which, given any $p$, is the probability that an arbitrary site of the diluted chain belongs to a cluster of $s$ number of sites \cite{stauffer_1992}. $n_s$ is the mean number of clusters of size $s$. This has a linear rise at small $p$ and a divergence near $p=1$. This analytic behavior, along with our numerical results, is shown in \Fig{fig2}(b)(here, $c_0=0.409 \pm 0.002$).  Broadly, the behavior indeed remains area law (with a $p$ dependent coefficient) except at $p=1$ where the logarithmic $L$ dependence is restored. Interestingly, the average cluster size $\langle M \rangle =\sum_s s P_s$, diverges near $p\rightarrow p_c$ as $(p_c-p)^{-\gamma}$ with $\gamma=1$ \cite{stauffer_1992}. Thus $\tilde{S}_{\text{quen}} \sim \frac{c}{6} \ln \langle M \rangle \sim \frac{c}{6} \ln \xi^{\frac{\gamma}{\nu}}$, where the geometric correlation length $\xi$ also diverges as $(p_c-p)^{-\nu}$ with $\nu =1$ \cite{stauffer_1992}. Since at $p=1$, $\xi \sim L$, one expects a scaling where, $ e^{6\tilde{S}_{\text{quen}}} L^{-\frac{\gamma}{\nu}}\sim 1$ which is shown in \Fig{fig2}(c). Thus, the entanglement measure captures the percolation exponents near the geometrical phase transition here at $p_c=1$. However, as discussed before, these results required us to put an infinitesimal degeneracy splitting disorder $\eta$, which also breaks the symmetry of the full Hamiltonian. Without any such disorder, the massive degenerate manifold of zero-modes leads to geometrical components of EE, as we discuss next. 

As the one-dimensional lattice is percolated, various clusters of different sizes appear on the chain. For any odd $s$ the cluster has one single particle zero energy mode. Thus, the zero-mode density is,
\beq 
\frac{{N}_0}{L} = \sum_{m=0}^{\infty} n_{2m+1} = \frac{1-p}{1+p}, \label{n0eq1d}
\eeq
for a $L$ sized chain at percolation probability $p$. This analytical behavior and the disorder averaged numerical estimate of zero-modes number $\tilde{N_0}$ are shown in \Fig{fig2}(e). A half-filled state in such a system again leads to highly degenerate many-body eigenspace. A uniform Haar measure here leads to 
\beq
\tilde{S}_{\text{typ}} = \tilde{S}_{\text{quen}} +  (\ln 2 -\frac{1}{2}) \tilde{N}_0,
\eeq
while $\tilde{S}_{\text{ann}} \sim \frac{\tilde{N}_0}{2} \ln2 $. All the behaviors match the numerical results as shown in \Fig{fig2}(d). Interestingly, the mutual information between the two subsystems removes this large volume law contribution and shows a rise similar to $\tilde{S}_{\text{quen}}$, only near $p=1$ when the lattice gets connected (see SM \cite{supp}). Thus, geometrical disorder, as in one-dimensional percolation, provides distinctive signatures in various entanglement measures both from intra-cluster fermionic correlations and from geometric components of the lattice itself.

\begin{figure}
\centering
\includegraphics[width=0.99\linewidth]{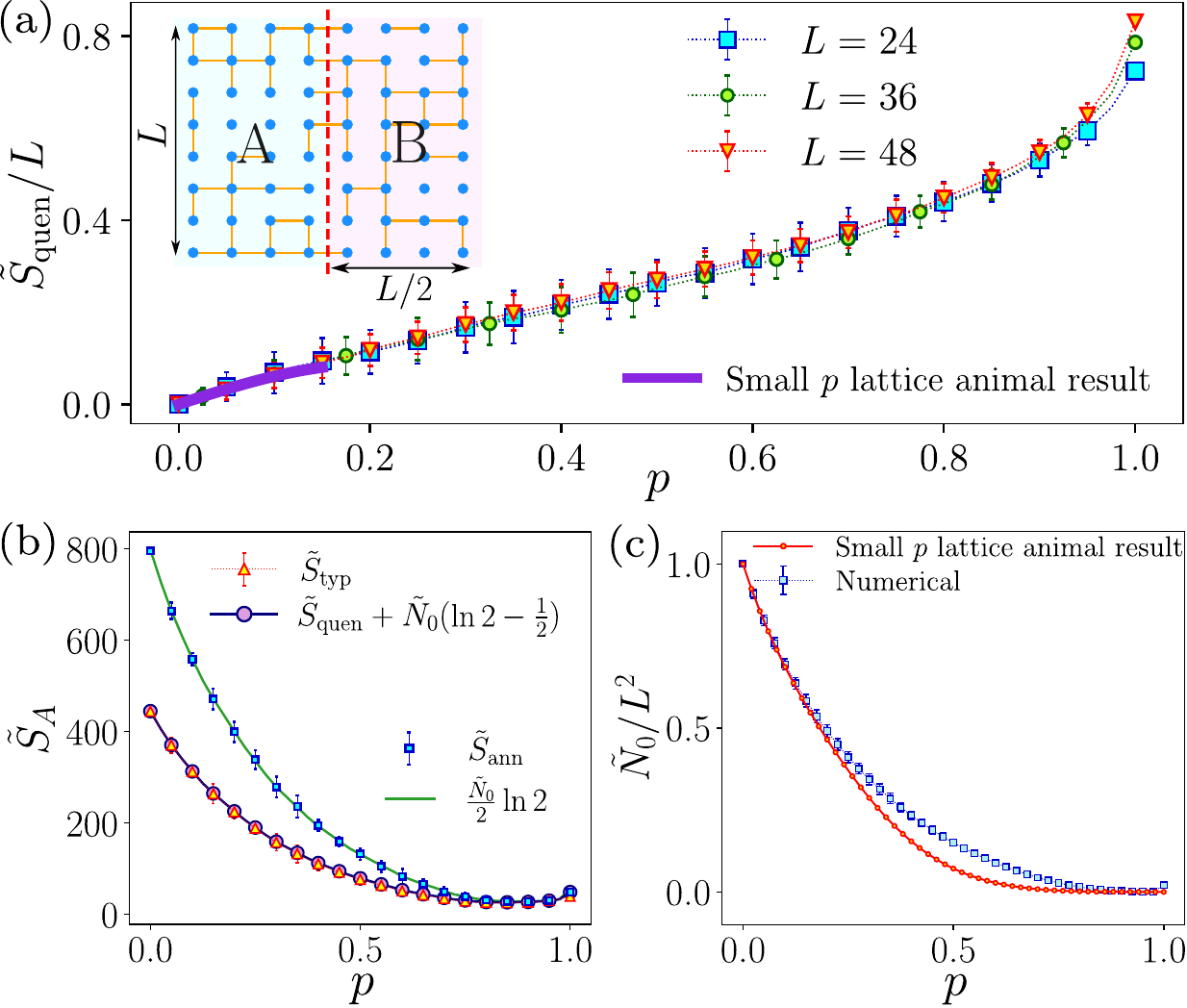}
	\caption{\textbf{Bond diluted square lattice:} (a) $\tilde{S}_{\text{quen}}$ with $p$ for different system sizes. ($N_C=10^2$). (inset) Schematic of a configuration with A and B partitions. (b)  $\tilde{S}_{\text{typ}}$ and $\tilde{S}_{\text{ann}}$ with $p$. $N_C=10, N_R=10^2$ (c) $\tilde{N}_0$  with $p$  ($N_C=10^2$), compared to lattice animal results (see text). In (b-c), $L=48$.}
	\label{fig3} 
\end{figure}

{\it \textbf{Two-dimensional percolation:}} In two dimensional percolation the system has a finite $p_c$; for instance in square lattice bond percolation it is known $p_c=\frac{1}{2}$ \cite{Sykes_PRL_1963, Sykes_JMP_1964,  Kesten_Springer_1980}. At $p=p_c$ a spanning cluster develops with critical exponent $\gamma=43/18$ \cite{stauffer_1992}. We again pose the question of different entanglement measures for this system. As is known that for a two-dimensional free fermionic system - any infinitesimal disorder localizes all the wavefunctions \cite{Anderson_PR_1958, Abrahams_PRL_1979}, thus in terms of electronic properties, we expect the system to be localized for all values of $p$ \cite{Kirkpatrick_PRB_1972, Raghavan_PRB_1981, Shapir_PRL_1982, Soukoulis_PRB_1991} even though there have been studies finding numerical evidence otherwise \cite{Srivastava_PRB_1984, Koslowski_PRB_1990, Nazareno_PRB_2002, Islam_PRE_2008, Schubert_PRB_2008, Dillon_EPJB_2014, Tomasi_PRB_2022}. Given any $p$ the mean number of clusters containing $k$ bonds, $n_k$ is given by $
n_k =\sum_t g(k,t)p^k(1-p)^t$ 
where $t$ is the perimeter of the cluster and $g(k,t)$ is the geometrical factor associated with the number of lattice animals given ($k,t$) \cite{Sykes_JPA_1981}. Since any square lattice is a bipartite graph with a symmetric spectrum, the $E_F$ remains pinned to zero at half-filling even under percolation. We find that $S_{\text{quen}}$ follows an area law behavior (see \Fig{fig3}(a)) i.e. $\propto L$ for $p<1$. While at $p=1$, the complete square lattice is restored, leading to a finite Fermi sea, the entanglement is  $\sim L\ln L$ given by the Widom conjecture \cite{Gioev_PRL_2006, Swingle_PRL_2010}. At $p<1$, however, such a momentum space description is no longer applicable. At small $p$, one can enumerate the lattice animals exactly and count their entanglement contribution, as shown by the analytical curve in \Fig{fig3}(a) (for details, see SM \cite{supp}). This is in contrast to $S_{\text{typ}}$ and $S_{\text{ann}}$, which again depend on the extensive number of zero-modes present in the system (see \Fig{fig3}(b)). The density of zero-modes can be estimated analytically from lattice animals, given by 
\beq
\frac{N_{0}}{L^2} = \sum_{k,t} n_0(k, t) g(k,t)p^k(1-p)^t,  \label{sqlatn0eqn}
\eeq 
where $n_0(k, t)$ is the number of zero-modes in a cluster of bond-size $k$ and perimeter $t$. A lower bound on $N_0$, calculated using lattice animals up to $k=4$, is shown in \Fig{fig3}(c), and it matches well with disorder averaged zero-mode number $\tilde{N}_0$ for small $p$ values (details in SM \cite{supp}). Given the zero-modes really appear from the geometrical aspects of the clusters, it is thus imperative that they also determine the entanglement content. 

\begin{figure}
\centering
\includegraphics[width=0.99\linewidth]{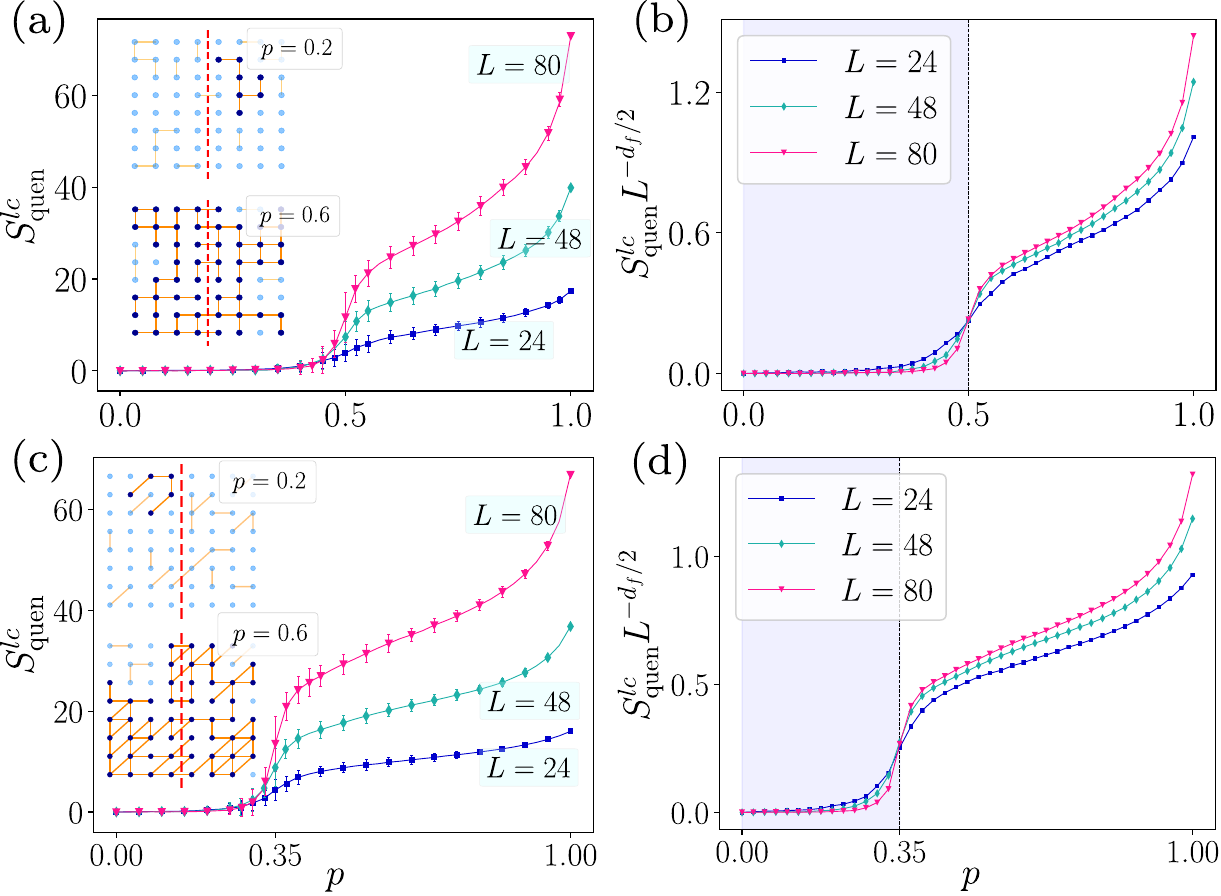}
	\caption{\textbf{Largest cluster:} (a) Disorder averaged $S^{lc}_{\text{quen}}$ with $p$, for square lattice of different size $L$  ($N_c=10^2$). (insets) typical configurations for two values of $p$ where {\it dark-blue} sites form the largest cluster. (b) Scaling of $S^{lc}_{\text{quen}}$ with the fractal dimension $d_f = 91/48$ shows crossing at percolation threshold $p_c=0.5$. In (c-d) similar to (a-b) but for triangular lattice with $p_c \sim 0.35$ (see text).}
	\label{fig4}
\end{figure}

To distill any signature of percolation transition at $p=p_c$, we calculate the configuration averaged quenched bipartite EE of the {\it largest} cluster ($\equiv S^{lc}_{\text{quen}}$) as a function of $p$ and show this in \Fig{fig4}(a). Interestingly, while for $p<p_c$  $S^{lc}_{\text{quen}} \propto L^0 $, for $p>p_c$ $S^{lc}_{\text{quen}} \propto L $ where $L$ is the linear dimension of the lattice. Given it is known that the largest cluster follows a scaling $L^{d_f}$ \cite{stauffer_1992} where $d_f$ is the fractal dimension of the system, for a typical area law behavior, we expect $S^{lc}_{\text{quen}} \propto L^{d_f/2}$ near $p_c$. A scaling collapse using this form shows a crossing (see \Fig{fig4}(b)) at $p=p_c=0.5$ with, $d_f=91/48$ as known for two-dimensional percolation transition. Since this physics should be independent of the microscopic lattice, we apply the same analysis to a tight-binding triangular lattice (\Fig{fig4}(c)). Again, the same collapse shows a crossing (\Fig{fig4}(d)) near $p=p_c=2\sin(\frac{\pi}{18})$ \cite{Sykes_JMP_1964} illustrating that the entanglement scaling indeed follows the universal features of geometrical phase transitions
even though the exact value of $p_c$ is itself not universal. 

{\it \textbf{Outlook:}} Quantum entanglement and its measures have taken a defining role in deciphering the nature of quantum phases. In particular, the nature of low-energy excitations are often equated with whether the bipartite EE follows area law or has logarithmic corrections. In this work, we revisit various measures of quantum entanglement in the context of percolation disorder in free fermionic lattice Hamiltonians. We find that percolation disorder inherently generates extensive degeneracies, which gives rise to subtleties in standard bipartite EE. It is then important to either break the massive degeneracies by putting infinitesimal disorder which leads to the $S_{\text{quen}}$, or otherwise investigate quantities such as $S_{\text{typ}}$ and $S_{\text{ann}}$ which includes physics of the degenerate manifold. We uncover that such measures have contributions from both fermionic correlations and geometrical aspects. While  $S_{\text{quen}}$ generically follow area law, $S_{\text{typ}}$  and $S_{\text{ann}}$ are volume law in character. These quantities can, in turn, be estimated from the properties of the clusters, which either cut the entanglement bipartition or contribute to the zero-mode degeneracies. Interestingly, the entanglement measure of the largest cluster can capture even the classical percolation threshold in two dimensions. While we have restricted our investigation to three quantities $S_{\text{typ}}, S_{\text{ann}}$ and $S_{\text{quen}}$, it would be worthwhile to quantify the amount of classical and quantum correlations in these systems. In physical systems where a perturbative quenched disorder is inherent, it is expected that $S_{\text{quen}}$ represents a more meaningfully observable quantity than $S_{\text{typ}}$. Similarly, given in estimation of $S_{\text{ann}}$ complete density matrix is averaged, the annealed EE contains both classical and quantum correlations. In this context, quantities such as mutual information and entanglement negativity may be of interest.

While in this work, we do not propose any experimental setups to measure such entanglement signatures, finding realistic proposals \cite{ Islam_Nature_2015, Kaufman_Science_2016, Lukin_Science_2019, Canella_SR_2019} in this direction would be interesting to pursue. Similarly, we have focused in this study exclusively on bipartite EE, given its relevance to quantum condensed matter systems. Various other measures have been pursued in quantum information to investigate phases and phase transitions \cite{zeng2019quantum, De_Chiara_2018}. A comprehensive investigation of these, in regard to percolation disorder, is another prospective study. Finally, investigating this physics in both symmetry-protected topological systems and topologically ordered systems will be an exciting future direction.

{\it \textbf{Acknowledgement:}} We acknowledge fruitful discussions with Diptarka Das, Apratim Kaviraj, Saikat Ghosh, G.~Sreejith, Saikat Mondal, and Harish Adsule. S.P. acknowledges funding from IIT Kanpur Institute Fellowship. AA acknowledges support from IITK Initiation Grant (IITK/PHY/2022010). Numerical calculations were performed on the workstation {\it Wigner} at IITK.

\bibliography{PercoEE_References}

\begin{thebibliography}{101}
\bibitem [{tol()}]{tol}%
  \BibitemOpen
  \href@noop {} {}\bibinfo {note} {Numerically, we always associate zero-modes
  with states lying in the energy gap $\Delta E\sim 10^{-14}$ around
  $E=0$.}\BibitemShut {Stop}%
\end{thebibliography}

\newpage
\setcounter{equation}{0}
\makeatletter
\renewcommand{\theequation}{S\arabic{equation}}
\renewcommand{\thefigure}{S\arabic{figure}}

\onecolumngrid

\begin{center}
	\textbf{\large Supplemental Material to ``Entanglement signatures of a percolating quantum system"}
\end{center}

\vspace{\columnsep}
\vspace{\columnsep}

\twocolumngrid

\section{Entanglement between two free spins} \label{2spin}

\begin{figure}
\centering
\includegraphics[width=0.95\linewidth]{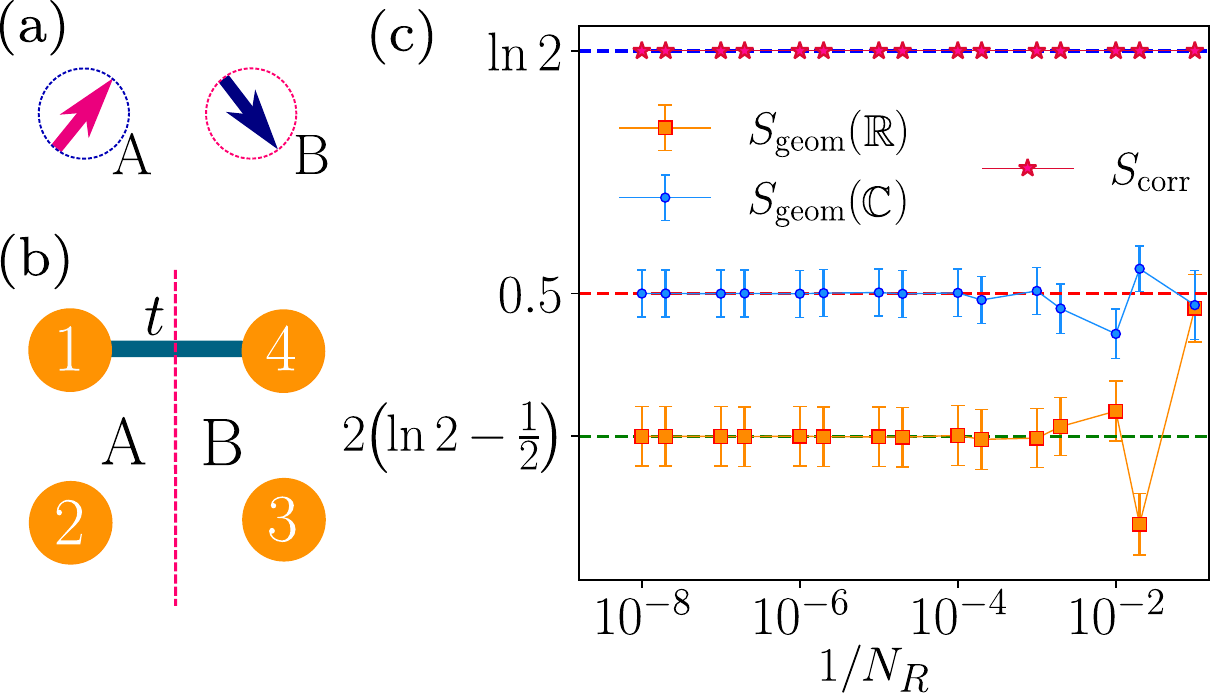}\caption{(a) Two free spin $A$ and $B$. (b) A four-site toy model with only two sites ($1$ and $4$) are connected via fermionic hopping of strength $t$. The bipartition used for entanglement calculation is shown in {\it red-dashed} line. (c) Different contributions in $S_{\text{typ}}$ for the four-site toy example. Contribution from fermionic correlation $S_{\text{corr}}$ is always $\ln 2$. For complex random coefficients, the geometric contribution $S_{\text{geom}}$ converges to $S_{\text{geom}} (\mathbb{C}) = \half$ with increasing $N_{R}$ (number of random pure state ensembles), and for real choices it converges to $S_{\text{geom}} (\mathbb{R}) = (2\ln 2 - 1)$.}
	\label{smfig1} 
\end{figure}

To understand the notion of {\it typical} and {\it annealed} entanglement in contrast with standard bipartite entanglement entropy, let us consider two free spins $A$ and, $B$ as shown in \Fig{smfig1}(a). A general choice of the valid many-body wavefunction of the system will be, 
\beq
\ket{\psi} = a\ket{ \downarrow_A \downarrow_B} +b\ket{ \downarrow_A \uparrow_B} +c\ket{ \uparrow_A \downarrow_B} +d\ket{ \uparrow_A \uparrow_B},
\eeq
 since the spins are free. Given the wavefunction coefficients (here $a, b, c$ and $d$) are complex random variables chosen from a Gaussian distribution with zero mean and unit variance, the random state $\ket{\psi}$ can be thought of as Haar uniform pure state \cite{Bengtsson_Book_2006, Nadal_JSP_2011}. So, in a general form, any random pure state of our free spin system can be written as,
\beq
\ket{\psi} = \sum_{i,j} \phi_{i,j}\ket{\alpha_{A}^{i}}\otimes\ket{\alpha_{B}^{j}},
\eeq
with $\alpha^i, \alpha^j \in \{\uparrow, \downarrow\}$ is the basis of the Hilbert space and wavefunction coefficients $\phi_{i,j}$s (i.e.~$a, b, c, d$) are the elements of a Gaussian random matrix $\Phi$, satisfying the wavefunction normalisation constraint $\sum_{i,j} |\phi_{i,j}|^2 = |a|^2+|b|^2+|c|^2+|d|^2 =1$. Now, the partial density matrix for the spin A will be,
\bea
\hat{\rho}_A &=& \text{Tr}_{B} (\hat{\rho}) = \text{Tr}_{B} \big(\ket{\psi}\bra{\psi}\big) \nonumber \\
&=& \sum_m \bra{\alpha_{B}^m} \bigg( \sum_{i,j} \sum_{k,l} \phi_{i,j} \phi_{k,l}^* \ket{\alpha_{A}^i}\bra{\alpha_{A}^k}  \otimes \ket{\alpha_{B}^j}\bra{\alpha_{B}^l} \bigg) \ket{\alpha_{B}^m} \nonumber  \\
&=& \sum_{i,k}\sum_m \phi_{i,m} \phi^{*}_{k,m}\ket{\alpha_{A}^i}\bra{\alpha_{A}^k} \nonumber  \\ 
&=& \sum_{i,k}\chi_{i,k} \ket{\alpha_{A}^i}\bra{\alpha_{A}^k},
\eea
where $\chi_{i,k} = \sum_m \phi_{i,m} \phi^{*}_{k,m}$ is the matrix element of $\Phi \Phi^{\dagger}$, consequently $\hat{\rho}_A = \Phi \Phi^{\dagger}$ in the basis of $\alpha_A$. From this, it is straightforward to see in our system, 
\beq
\hat{\rho}_A = \begin{pmatrix}
|a|^2+|b|^2 & ac^* + bd^* \\ ca^* + db^* & |c|^2+|d|^2
\end{pmatrix}.
\eeq
Now using, $\hat{\rho}_A$ one can numerically calculate bipartite entanglement entropy $S_A (\hat{\rho}_A) = \text{Tr} \big( {\hat{\rho}_A \ln \hat{\rho}_A} \big)$ and average it over numerous Gaussian ensembles of wavefunction coefficients \{$a, b, c, d$\} to find {\it typical} entanglement measure $S_{\text{typ}} = \langle S_A(\hat{\rho}_A) \rangle \simeq 0.33$. We also define {\it annealed} entanglement as $S_{\text{ann}} = S_A(\langle \hat{\rho}_A \rangle) = \text{Tr} \big( {\langle \hat{\rho}_A \rangle \ln \langle \hat{\rho}_A \rangle} \big)$. To calculate this, one first takes ensemble averaged (this average is also taken over wavefunction coefficients as described earlier) partial density matrix $\langle \hat{\rho}_A \rangle$ which turns out to be $\half I_2$ here, and then find entanglement entropy to get $S_{\text{ann}}=\ln2$.

\section{Four site fermionic toy example} \label{4site}

To illustrate how {\it typicality} in entanglement measure appears in a fermionic model, consider the system where two spinless fermions hops (with hopping strength $t=-1$) on a four-site structure as shown in \Fig{smfig1}(b). While the single-particle spectrum is straight-forwardly given by $\{\pm 1,0,0\}$, the many-body ground state is not unique but two-fold degenerate. This can be seen because, to minimize the energy, while one fermion resides in a delocalized state between sites 1 and 4, another fermion can be on either site 2 or 3. The two valid ground-state wavefunctions of the system are  
\bea
|\Psi_1\rangle &=& \frac{1}{\sqrt{2}} ( c^\dagger_1 + c^\dagger_4 ) c^\dagger_2 | \Omega\rangle, \\
|\Psi_2\rangle &=& \frac{1}{\sqrt{2}} ( c^\dagger_1 + c^\dagger_4 ) c^\dagger_3 | \Omega\rangle,
\eea
where $|\Omega\rangle$ is the fermionic vacuum and the 
entanglement entropy across the partition is clearly $\ln 2$ given the delocalized fermion. However, any general wavefunction of the form 
\beq
|\Psi \rangle = \phi_1|\Psi_1 \rangle + \phi_2 | \Psi_2 \rangle, 
\eeq
is also a valid wave function of the system. Given the coefficients are randomly drawn from a Gaussian distribution and follow wavefunction normalization constraint, they form a random pure state similar to what we discussed in \ref{2spin}. This leads to similar considerations of {\it typical} entanglement $S_{\text{typ}}$ as described in the main text.
However, now it has components arising from the fermionic correlations within the connected cluster across the partition (the delocalized state between site 1 and 4) and contributions due to a random choice of coefficients $\phi_{1}$ and $\phi_2$. In general, 
\beq
S_{\text{typ}} = S_{\text{corr}} + S_{\text{geom}},
\eeq
where $S_{\text{corr}}$ captures the fermionic correlations in connected regions and $S_{\text{geom}}$ arises out of typicality due to degeneracy, often arising out of disconnected sites. Therefore, the latter has information about the geometrical properties of the network. In the present example, while $S_\text{corr} = \ln2$, $S_{\text{geom}}$ depends on the choices of the wavefunction coefficients ($\phi_1$ and $\phi_2$) used in the formation of random pure states, leading to the symmetry classification of the typical entanglement \cite{Liu_PRB_2023}. In class A \cite{altland_1997}, the coefficients are complex, and the analytical result shows $S_{\text{geom}}(\mathbb{C})=\half$,  while in class AI \cite{altland_1997} where the coefficients are real, it is evaluated to be $S_{\text{geom}}(\mathbb{R})=(2\ln 2 - 1)$. In \Fig{smfig1}(c), we show that our numerical results converge towards analytical results with the increasing number of pure ensembles ($N_{R}$) used in the calculation of $S_{\text{typ}}$.

\begin{figure}
\centering
\includegraphics[width=0.92\linewidth]{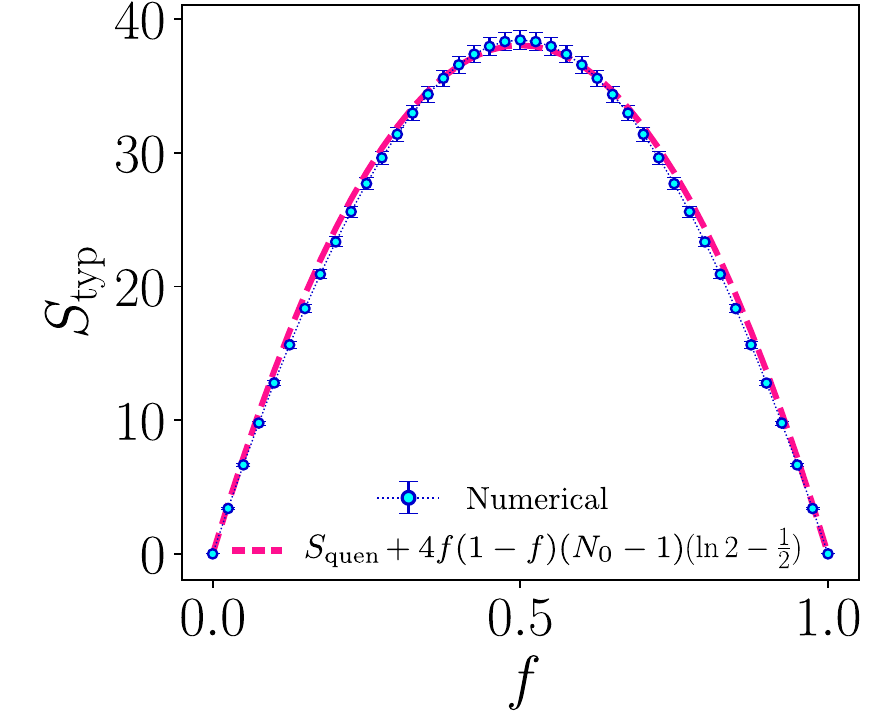}
\caption{$S_{\text{typ}}$ as a function of filling $f$, for the zero-dimensional fermionic problem. Numerical data is calculated for a system of $N=200$ sites with equal bipartition and averaged over $10^4$ ensembles of random pure states.}
\label{smfig2} 
\end{figure}

\section{Zero-dimensional fermionic problem} \label{0dtb}

An all-to-all connected fermionic hopping network hosts $N_0 = N-1$ number of zero energy states for a $N$ site system. So, a typical many-body wavefunction of the system will be a random pure state made out of these degenerate zero energy states. Here, we choose the wavefunction coefficients to be real, restricting it to class AI \cite{Liu_PRB_2023}. Now, $S_{\text{typ}}$ at half filling $f=\half$, and for equal bipartition, typical entanglement contains a small fermionic correlation part and a substantial geometric contribution because of the massive degeneracy in the spectrum. While the geometric part follows the symmetry classification result \cite{Liu_PRB_2023}, the fermionic part is captured in $S_{\text{quen}}$, evaluated using the degeneracy breaking protocol as discussed in the main text. Now, for any general filling $f$ ($0<f<1$) with equal bipartition, the available phase space volume is $f(1-f)$; consequently, the effective degenerate Hilbert space dimension becomes $\sim 2^{af(1-f)N_0}$, where $a$ is a dimensionless constant. Then, using Page's result \cite{Page_PRL_1993}, we empirically find the typical entanglement in the leading order to be,
\beq
S_{\text{typ}}(f) = S_{\text{quen}}(f) + 4f\big(1-f\big)\big(N_0 - 1\big)\big(\ln 2 - \half \big), 
\eeq
so that it remains consistent with the result at half filling ($S_{\text{typ}} = S_{\text{quen}} + \big(N_0 - 1\big)\big(\ln 2 - \half \big)$). In \Fig{smfig2}, we show the numerical values as a function of filling ($f$), which match well with our empirical formula.

\section{Mutual information in one-dimensional percolation}
\begin{figure}
\centering
\includegraphics[width=0.92\linewidth]{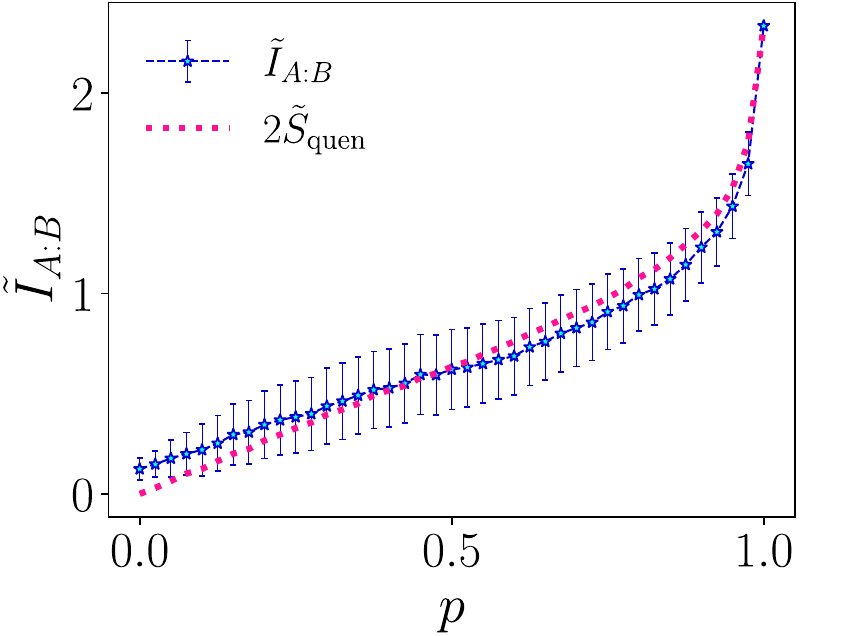}
\caption{Disorder averaged MI $\tilde{I}_{A:B}$ with percolation probability $p$ for the free fermionic chain of length $L=200$. The bipartition protocol is the same as the main text. The disorder average is taken over $500$ configurations, and for each configuration, $10^3$ random pure states are chosen for averaging the correlator matrix. Note the similarity in behaviour with $\tilde{S}_{\text{quen}}$ (same data as shown in the main text for $L=200$).}
	\label{smfig3} 
\end{figure}

In general, mutual information (MI) is a good measure of entanglement between two subsystems $A$ and $B$ when the full system is in mixed state \cite{Adami_PRA_1997}. Since finding annealed entanglement entropy ($S_{\text{ann}}$) in our study is associated with the mixedness of the full system, here we intend to show the behavior of MI in the percolated tight-binding chain. With the same bipartition protocol used in the main text, MI between $A$ and $B$ is, 
\beq
I_{A:B} = S^{A}_{\text{ann}} + S^{B}_{\text{ann}} - S^{AB}_{\text{ann}},
\eeq
where, the annealed entanglement entropy of the full system $S^{AB}_{\text{ann}} \neq 0$ due to mixed character of the correlator matrix for the full system at half filling. In \Fig{smfig3}, the behavior of disorder averaged MI ($\tilde{I}_{A:B}$) with bond occupation probability $p$ is shown for our one-dimensional percolation model. In $p\rightarrow0$ limit, $S^{A}_{\text{ann}} + S^{B}_{\text{ann}} \approx S^{AB}_{\text{ann}}$. While the behavior shows reduction of large volume law (due to zeromodes) contribution in $S_{\text{ann}}$ for small $p$ values, a sharp rise appears near the classical percolation threshold $p = p_c = 1$, where the lattice gets connected. This is similar to  $\tilde{S}_{\text{quen}}$ (see \Fig{smfig3}), which is expected because, after the removal of zeromode contribution, only the electronic correlation part reflects in MI. 

\section{Lattice animals, their entanglement and number of zeromodes} \label{sqlat}

\begin{table}[htbp]
    \centering
    \resizebox{0.99\linewidth}{!}{
     \begin{tabular}{c|c|c|c|c|c}
        \hline
        \hline
         \begin{tikzpicture}[scale=0.2]
        \draw[white, ultra thin] (0.5,-0.6) -- (0.5,1.1);
        \end{tikzpicture}
        \textbf{Animals} & \bm{$S_{\textbf{animal}}^{b}$} & \bm{$P_{\textbf{animal}}^{b}$} & 
        \bm{$k, t$} &
        {$\bm{g(k,t)}$} &
        \bm{$n_0(k,t)$} \\
        \hline
        \hline
        \begin{tikzpicture}[scale=0.2]
        \draw[white, ultra thin] (0,-0.6) -- (0,1.3);
            \filldraw[red, very thick](0,0) circle (6pt);
        \end{tikzpicture}&
        $0$&
        $q^4$&
        $0, 4$&
        $1$&
        $1$\\
        \hline
        \begin{tikzpicture}[scale=0.2]
        \draw[white, ultra thin] (0,-1.0) -- (0,1.5);
            \draw[blue, very thick] (-1,0) -- (1,0);
            \filldraw[red, very thick](-1,0) circle (6pt);
            \filldraw[red, very thick](1,0) circle (6pt);
            \draw[black, dotted, very thick] (0,-1.1) -- (0,1.1);
        \end{tikzpicture}
        & 
        $\ln 2$
        &
        $pq^6$&
        $1, 6$&
        $2$&
        $0$\\
        \hline
        \begin{tikzpicture}[scale=0.2]
            \draw[white, ultra thin] (0,-1.5) -- (0,1.7);
            \draw[blue, very thick] (-1,-0.5) -- (1,-0.5);
            \draw[blue, very thick] (-1,-0.5) -- (-1,1.2);
            \filldraw[red, very thick](-1,-0.5) circle (6pt);
            \filldraw[red, very thick](-1,1.2) circle (6pt);
            \filldraw[red, very thick](1,-0.5) circle (6pt);
            \draw[black, dotted, very thick] (0,-1.5) -- (0,1.0);
        \end{tikzpicture}
        & 
        $0.56234...$
        &
        $6p^2q^8$&
        $2, 8$&
        $6$&
        $1$\\
        \hline
        \begin{tikzpicture}[scale=0.2]
         \draw[white, ultra thin] (0,-1.0) -- (0,0.8);
            \draw[blue, very thick] (-1,-0.5) -- (1,-0.5);
            \draw[blue, very thick] (1,-0.5) -- (1,-2.2);
             \draw[blue, very thick] (1,-2.2) -- (3,-2.2);
            \filldraw[red, very thick](-1,-0.5) circle (6pt);
            \filldraw[red, very thick](1,-2.2) circle (6pt);
            \filldraw[red, very thick](3,-2.2) circle (6pt);
            \filldraw[red, very thick](1,-0.5) circle (6pt);
            \draw[black, dotted, very thick] (0,-1.7) -- (0,0.6);
        \end{tikzpicture}
        & 
        $\ln 2$
        &
        $14p^3q^{10}$& & &\\
        \cline{1-3} 
        \begin{tikzpicture}[scale=0.2]
         \draw[white, ultra thin] (0,-1.5) -- (0,1.8);
            \draw[blue, very thick] (-1,-0.5) -- (1,-0.5);
            \draw[blue, very thick] (-1,-0.5) -- (-1,1.2);
            \draw[blue, very thick] (1,-0.5) -- (3,-0.5);
            \filldraw[red, very thick](-1,-0.5) circle (6pt);
            \filldraw[red, very thick](-1,1.2) circle (6pt);
            \filldraw[red, very thick](1,-0.5) circle (6pt);
            \filldraw[red, very thick](3,-0.5) circle (6pt);
            \draw[black, dotted, very thick] (0,-1.5) -- (0,1.0);
        \end{tikzpicture}
        & 
        $0.41328...$
        &
        $p^3(7q^{10} + 2q^9)$&
        $3, 10$&
        $14$&
        $0$\\
        \hline
        \begin{tikzpicture}[scale=0.2]
            \draw[white, ultra thin] (0,-1.5) -- (0,1.8);
            \draw[blue, very thick] (-1,-0.5) -- (1,-0.5);
            \draw[blue, very thick] (-1,-0.5) -- (-1,1.2);
            \draw[blue, very thick] (-1,-0.5) -- (-3,-0.5);
            \filldraw[red, very thick](-1,-0.5) circle (6pt);
            \filldraw[red, very thick](-1,1.2) circle (6pt);
            \filldraw[red, very thick](1,-0.5) circle (6pt);
            \filldraw[red, very thick](-3,-0.5) circle (6pt);
            \draw[black, dotted, very thick] (0,-1.5) -- (0,1.0);
        \end{tikzpicture}
        & 
        $0.45056...$
        &
        $6p^3q^{10}$&
        $3,10$&
        $4$&
        $1$\\
        \hline
        \begin{tikzpicture}[scale=0.2]
            \draw[blue, very thick] (-1,-0.5) -- (1,-0.5);
            \draw[blue, very thick] (1,-0.5) -- (1,-2.2);
             \draw[blue, very thick] (1,-2.2) -- (-1,-2.2);
            \filldraw[red, very thick](-1,-0.5) circle (6pt);
            \filldraw[red, very thick](1,-2.2) circle (6pt);
            \filldraw[red, very thick](-1,-2.2) circle (6pt);
            \filldraw[red, very thick](1,-0.5) circle (6pt);
            \draw[black, dotted, very thick] (0,-2.7) -- (0,0.2);
        \end{tikzpicture}
        & 
        $1.17903...$
        &
        $2p^3q^9$&
        $3,9$&
        $4$&
        $0$ \\
        \hline
        \begin{tikzpicture}[scale=0.2]
         \draw[white, ultra thin] (0,-1.5) -- (0,1.8);
            \draw[blue, very thick] (-1,-0.5) -- (1,-0.5);
            \draw[blue, very thick] (-1,-0.5) -- (-1,1.2);
            \draw[blue, very thick] (3,-0.5) -- (3,1.2);
            \draw[blue, very thick] (1,-0.5) -- (3,-0.5);
            \filldraw[red, very thick](-1,-0.5) circle (6pt);
            \filldraw[red, very thick](-1,1.2) circle (6pt);
            \filldraw[red, very thick](1,-0.5) circle (6pt);
            \filldraw[red, very thick](3,-0.5) circle (6pt);
            \filldraw[red, very thick](3,1.1) circle (6pt);
            \draw[black, dotted, very thick] (0,-1.5) -- (0,1.0);
        \end{tikzpicture}
        & 
        $0.54387...$
        &
        $p^4(34q^{12}+8q^{11})$ & & &
        \\
        \cline{1-3}
        \begin{tikzpicture}[scale=0.2]
         \draw[white, ultra thin] (0,-1.5) -- (0,1.7);
            \draw[blue, very thick] (-1,-0.5) -- (1,-0.5);
            \draw[blue, very thick] (-1,-0.5) -- (-1,1.1);
            \draw[blue, very thick] (-1,-0.5) -- (-1,-2.1);
            \draw[blue, very thick] (-1,-2.1) -- (-3,-2.1);
            \filldraw[red, very thick](-1,-0.5) circle (6pt);\filldraw[red, very thick](-3,-2.1) circle (6pt);
            \filldraw[red, very thick](-1,1.1) circle (6pt);
            \filldraw[red, very thick](1,-0.5) circle (6pt);
            \filldraw[red, very thick](-1,-2.1) circle (6pt);
            \draw[black, dotted, very thick] (0,-1.5) -- (0,1.0);
        \end{tikzpicture}
        & 
        $0.56234...$
        &
        $p^4(20q^{12}+8q^{11})$& & &
        \\
        \cline{1-3}
        \begin{tikzpicture}[scale=0.2]
            \draw[white, ultra thin] (0,-1.5) -- (0,1.7);
            \draw[blue, very thick] (-1,-0.5) -- (1,-0.5);
            \draw[blue, very thick] (1,1.1) -- (1,-0.5);
            \draw[blue, very thick] (1,-0.5) -- (1,-2.1);
            \draw[blue, very thick] (-1,-0.5) -- (-3,-0.5);
            \filldraw[red, very thick](-1,-0.5) circle (6pt);
            \filldraw[red, very thick](1,-2.1) circle (6pt);
            \filldraw[red, very thick](1,-0.5) circle (6pt);
            \filldraw[red, very thick](-3,-0.5) circle (6pt);
            \filldraw[red, very thick](1,1.1) circle (6pt);
            \draw[black, dotted, very thick] (-2,-1.5) -- (-2,1.0);
        \end{tikzpicture}
        & 
        $\ln 2$
        &
        $6p^4q^{12}$& & &
        \\
        \cline{1-3}
        \begin{tikzpicture}[scale=0.2]
            \draw[white, ultra thin] (0,-0.0) -- (0,2.0);
            \draw[blue, very thick] (-1,-2.1) -- (1,-2.1);
            \draw[blue, very thick] (1,1.1) -- (1,-0.5);
            \draw[blue, very thick] (1,-0.5) -- (1,-2.1);
            \draw[blue, very thick] (-1,1.1) -- (1,1.1);
            \filldraw[red, very thick](1,-0.5) circle (6pt);
            \filldraw[red, very thick](1,-2.1) circle (6pt);
            \filldraw[red, very thick](-1,1.1) circle (6pt);
            \filldraw[red, very thick](-1,-2.1) circle (6pt);
            \filldraw[red, very thick](1,1.1) circle (6pt);
            \draw[black, dotted, very thick] (0,-2.5) -- (0,1.8);
        \end{tikzpicture}& 
        $1.14317...$&
        $2p^4q^{12}$&
        $4, 12$&
        $54$&
        $1$\\
        \hline
        \begin{tikzpicture}[scale=0.2]
         \draw[white, ultra thin] (0,-0.5) -- (0,1.0);
            \draw[blue, very thick] (-1,-0.5) -- (1,-0.5);
            \draw[blue, very thick] (-1,-2.2) -- (-1,-0.5);
            \draw[blue, very thick] (-1,-0.5) -- (-3,-0.5);
            \draw[blue, very thick] (-1,-2.2) -- (1,-2.2);
            \filldraw[red, very thick](-1,-0.5) circle (6pt);\filldraw[red, very thick](1,-2.2) circle (6pt);
            \filldraw[red, very thick](-3,-0.5) circle (6pt);
            \filldraw[red, very thick](1,-0.5) circle (6pt);
            \filldraw[red, very thick](-1,-2.2) circle (6pt);
            \draw[black, dotted, very thick] (0,-3.0) -- (0,0.7);
        \end{tikzpicture}
        & 
        $1.17349...$
        &
        $8p^4q^{11}$ & & &
        \\
        \cline{1-3}
        \begin{tikzpicture}[scale=0.2]
         \draw[white, ultra thin] (0,-0.5) -- (0,1.0);
            \draw[blue, very thick] (-1,-0.5) -- (1,-0.5);
            \draw[blue, very thick] (-1,-2.2) -- (-1,-0.5);
            \draw[blue, very thick] (1,-0.5) -- (3,-0.5);
            \draw[blue, very thick] (-1,-2.2) -- (1,-2.2);
            \filldraw[red, very thick](-1,-0.5) circle (6pt);\filldraw[red, very thick](1,-2.2) circle (6pt);
            \filldraw[red, very thick](3,-0.5) circle (6pt);
            \filldraw[red, very thick](1,-0.5) circle (6pt);
            \filldraw[red, very thick](-1,-2.2) circle (6pt);
            \draw[black, dotted, very thick] (0,-3.0) -- (0,0.7);
        \end{tikzpicture}
        & 
        $0.95494...$
        &
        $8p^4q^{11}$ & & &
        \\
        \cline{1-3}
        \begin{tikzpicture}[scale=0.2]
            \draw[white, ultra thin] (0,-1.5) -- (0,2);
            \draw[blue, very thick] (-1,-0.5) -- (1,-0.5);
            \draw[blue, very thick] (1,1.1) -- (1,-0.5);
            \draw[blue, very thick] (-1,-0.5) -- (-1,1.2);
            \draw[blue, very thick] (-1,-0.5) -- (-3,-0.5);
            \filldraw[red, very thick](-1,-0.5) circle (6pt);
            \filldraw[red, very thick](-1,1.2) circle (6pt);
            \filldraw[red, very thick](1,-0.5) circle (6pt);
            \filldraw[red, very thick](-3,-0.5) circle (6pt);
            \filldraw[red, very thick](1,1.2) circle (6pt);
            \draw[black, dotted, very thick] (0,-1.5) -- (0,1.0);
        \end{tikzpicture}& 
        $0.32346...$ &
        $p^4(10q^{12}+8q^{11})$ & & &
        \\
        \cline{1-3}
        \begin{tikzpicture}[scale=0.2]
         \draw[white, ultra thin] (0,-1.0) -- (0,0.8);
            \draw[blue, very thick] (-1,-0.5) -- (1,-0.5);
            \draw[blue, very thick] (1,-0.5) -- (1,-2.2);
            \draw[blue, very thick] (3,-0.5) -- (3,-2.2);
             \draw[blue, very thick] (1,-2.2) -- (3,-2.2);
            \filldraw[red, very thick](-1,-0.5) circle (6pt);
            \filldraw[red, very thick](1,-2.2) circle (6pt);
            \filldraw[red, very thick](3,-2.2) circle (6pt);
            \filldraw[red, very thick](1,-0.5) circle (6pt);
            \filldraw[red, very thick](3,-0.5) circle (6pt);
            \draw[black, dotted, very thick] (0,-1.7) -- (0,0.6);
        \end{tikzpicture} & 
        $0.63651...$ &
        $p^4(30q^{12}+8q^{11})$ &
        $4,11$ &
        $32$ &
        $1$ \\
        \hline
        \begin{tikzpicture}[scale=0.2]
            \draw[white, ultra thin] (0,-0.5) -- (0,1.6);
            \draw[blue, very thick] (-1,-0.5) -- (1,-0.5);
            \draw[blue, very thick] (-1,-0.5) -- (-1,1.1);
            \draw[blue, very thick] (-1,-0.5) -- (-1,-2.1);
            \draw[blue, very thick] (-1,-0.5) -- (-3,-0.5);
            \filldraw[red, very thick](-1,-0.5) circle (6pt);\filldraw[red, very thick](-3,-0.5) circle (6pt);
            \filldraw[red, very thick](-1,1.1) circle (6pt);
            \filldraw[red, very thick](1,-0.5) circle (6pt);
            \filldraw[red, very thick](-1,-2.1) circle (6pt);
            \draw[black, dotted, very thick] (0,-1.5) -- (0,1.0);
        \end{tikzpicture}
        & 
        $0.37677...$
        &
        $2p^4q^{12}$&
        $4,12$&
        $1$&
        $3$ \\
        \hline
        \begin{tikzpicture}[scale=0.2]
            \draw[white, ultra thin] (0,-0.5) -- (0,3.2);
            \draw[blue, very thick] (0,0) rectangle (2,2);
            \filldraw[red, very thick](2,2) circle (6pt);
            \filldraw[red, very thick](0,0) circle (6pt);
            \filldraw[red, very thick](2,0) circle (6pt);
            \filldraw[red, very thick](0,2) circle (6pt);
            \draw[black, dotted, very thick] (1,2.8) -- (1,-0.8);
        \end{tikzpicture}
        & 
        $0.83299...$
        &
        $p^4q^8$&
        $4,8$&
        $1$&
        $2$\\
        \hline
        \hline
    \end{tabular}%
    }
    \caption{Small-sized lattice animals (up to bond-size $k=4$ and perimeter $t=12$) and their contribution in quenched entanglement $S_{\text{quen}}$ and number of zeromodes $N_0$ for percolated square lattice. $S_{\text{animal}}^{b}$ is the entanglement of a lattice animal given the bipartition cuts the animal in a certain way, and $P_{\text{animal}}^{b}$ is the probability of that animal appears on that bipartition. The prefactors in $P_{\text{animal}}^{b}$ are decided by the number of animal structures with the same partition, giving the same $S_{\text{animal}}^{b}$. For a given \{$k, t$\}, $g(k,t)$ is the number of lattice animals, and $n_0(k,t)$ is the number of zero energy states they can host.}
    \label{tab:lat_ann}
\end{table}

In the case of square lattice percolation, the quenched entanglement $S_{\text{quen}}$ can be captured using the lattice animals and their contributions to entanglement given a bipartition. For a lattice animal of bond size $k$ (i.e., the number of bonds) and perimeter $t$, we calculate the bipartite entanglement entropy $S_{\text{animal}}^{b}$ with the partition dissecting the animal in a certain way. Now if $P_{\text{animal}}^{b}$ is the probability that the different animals of the same \{$k, t$\} appear on that given partition and give rise to the same $S_{\text{animal}}^{b}$, then for a square lattice of linear size $L$, 
\beq
S_{\text{quen}} \approx \sum_{\text{animal},b} P_{\text{animal}}^{b} S_{\text{animal}}^{b} L.
\label{sanimaleqn}
\eeq
The factor of $L$ comes because there are $L$ number of sites on the partition where the animal can appear. 

Similarly, we calculate the number of zero energy modes in the percolating square lattice using lattice animal enumeration. The mean number of bond clusters is given by $g(k,t)p^k(1-p)^t$ and $g(k,t)$ is the number of lattice animals of size $\{k, t\}$ \cite{Sykes_JPA_1981}. Then the number of zeromodes for $L$ sized system is, 
\beq
N_{0}= \sum_{k,t} n_0(k, t) g(k,t)p^k(1-p)^t L^2,  \label{sqlatn0eqn}
\eeq 
where $n_0(k, t)$ is the number of zeromodes in a cluster of size $\{k, t\}$. For small values of $k$ (up to $4$) and $t$ (up to $12$), all these data are tabulated in TABLE \ref{tab:lat_ann}. Now, using \eqn{sanimaleqn} and \eqn{sqlatn0eqn}, one can calculate quenched entanglement and the number of zeromodes in the percolated square lattices. Since these small-sized clusters are dominant in small values of percolation probability $p$, our result from TABLE \ref{tab:lat_ann}
matches the disorder averaged numerical results ($\tilde{S}_{\text{quen}}$ and $\tilde{N}_0$) at small $p$ values (see main text).

\end{document}